\documentclass[aps,pra,twocolumn,superscriptaddress,showpacs,showkeys,amsmath,amssymb]{revtex4}

\usepackage{amsfonts}
\usepackage{amssymb,amsmath}
\usepackage{mathrsfs}
\usepackage{latexsym}
\usepackage{amsmath}
\usepackage[cp1251]{inputenc}
\usepackage{graphicx}
\usepackage{dcolumn}
\usepackage{bm}
\usepackage{color}

\RequirePackage{ifthen}
\RequirePackage[pdfstartview=FitH]{hyperref}
\begin{document}
	
	\title{Universal $p$-wave tetramers in low-dimensional fermionic systems\\ with three-body interaction}
	\author{V.~Polkanov\footnote{e-mail: cogersum92@gmail.com}}
	\affiliation{Professor Ivan Vakarchuk Department for Theoretical Physics, Ivan Franko National University of Lviv, 12 Drahomanov Street, Lviv, Ukraine}
	\author{V.~Pastukhov\footnote{e-mail: volodyapastukhov@gmail.com}}
	\affiliation{Professor Ivan Vakarchuk Department for Theoretical Physics, Ivan Franko National University of Lviv, 12 Drahomanov Street, Lviv, Ukraine}

	\date{\today}

	
	
	\begin{abstract}
		Inspired by the narrow Feshbach resonance in systems with the two-body interaction, we propose the two-channel model of three-component fermions with the three-body interaction that takes into account the finite-range effects in low dimensions. Within this model, the $p$-wave Efimov-like effect in the four-body sector is predicted in fractional dimensions above 1D. The impact of the finite-range interaction on the formation of the four-body bound states in $d=1$ is also discussed in detail.
	\end{abstract}
	
	\maketitle
\section{Introduction}
A few-body quantum physics is known to be drastically different from its classical counterpart. The most famous example is the Efimov effect \cite{Efimov_70, Naidon_Endo_17, Greene_et_al} realizing, as an infinite tower of three-body bound states, in the three-boson system at resonant two-body interaction. Similar behavior is found \cite{Efimov_73, Petrov_03} in the three-body sector of two-component fermions. In the fermionic case, the Efimov effect occurs in the odd-wave (typically $p$-wave) channels on the arbitrary intervals of particle mass ratios. Being well-established in three dimensions, this effect, in general, has a limitation on spacial dimensionality \cite{Nielsen_01}. Even for non-equal particle masses the universal three-body bound states emerge in the non-integer dimensions between $2<d<4$. Exactly in 2D, the super Efimov effect was found \cite{Nishida_etal_13} in a system of the spin-polarized fermions with resonant $p$-wave interaction. Attempt to find universal Efimov-like behavior in lower dimensions necessarily requires the mixed-dimension geometries \cite{Nishida_11}, or higher-order few-body interactions, namely, four-body \cite{Nishida_Son_10} in 1D and three-body \cite{Nishida_17} in 2D. In the latter cases, the system parameters should be highly fine-tuned to provide the resonant highest-order interaction and vanishing all lower-order couplings.

In recent years, research in this field has shifted to fractional dimensions. Particularly, interesting analytical results for a model with zero-range two-body potential were obtained in Refs.~\cite{Mohapatra_18, Rosa_22, Garrido_22}. The existence of the Efimov-like bound states with characteristic scaling low in the four-body sector was reported in \cite{Hryhorchak_22} for a bosonic system with the resonant three-body interaction in non-integer ($1<d<2$) dimensions. Independently of the spatial dimension, for particles interacting through the realistic two-body potential, the three-body forces typically appear \cite{Valiente_19, Pricoupenko_Petrov_19, Valiente_Zinner} as the effective induced interaction. The latter becomes dominant \cite{Pricoupenko_Petrov_21} when the two-body potential is tuned to a zero crossing. For the effective field theories above the upper critical dimension, the introduction of three-body terms in the Lagrangians is mandatory to provide the UV completeness of theories. In context of the Efimov physics, such a completion was proposed in seminal papers by Bedaque et al. \cite{BHvK_99_1, BHvK_99_2}, and more recently in Ref.~\cite{Sekino_Nishida_21} for a model of 1D spin-polarized fermions with the contact two-body $p$-wave interaction.
A start for active investigation of 1D few- and many-body systems with three-body interaction should be identified with the publication of several articles \cite{Drut_18, Nishida_18, Guijarro_et_al, Pricoupenko_18}, describing general properties of $SU(3)$ fermions and (mostly) three-body states of bosons. Similarly to 2D systems of particles interacting with each other by a two-body $\delta$-(pseudo)potential \cite{Jackiw_91}, the three-body contact interaction exhibits the quantum scale anomaly in 1D, which predetermines universal properties and thermodynamics of diluted bosons \cite{Pastukhov_19, Valiente_Pastukhov} and fermions \cite{Maki_19, McKenney_20}. In the thermodynamic limit, the three-body interaction is responsible for the formation of the quantum droplet state \cite{Sekino_18, Morera_22} in the 1D system of bosons and for the crossover transition \cite{Tajima_22} from tightly bound vacuum trimers to the so-called Cooper triples. 

The present work deals with the two-channel model of three-component fermionic particles with unequal masses that interact through the three-body forces in fractional dimension. The simplified version of the proposed setup in harmonic trapping potential was studied in \cite{Czejdo_20} using the high-temperature series at finite densities of constituents. In particular, we are interested in finding regions in the parameter space for the emergence of the four-body bound states and revealing their peculiarities. Previously, important aspects of a few-body physics of $SU(3)$ fermions with the contact three-body interaction in 1D were discussed by McKenney and Drut in Ref.~\cite{McKenney_19}. In this context, we fully complement the four-body sector by revealing the bound states for a system with unequal masses of different fermionic species and generalize these results to a case of finite interaction ranges and higher dimensions.

\section{Model and renormalization}
We consider a model of three-component Galilean-invariant fermions with different masses $m_{\sigma}$ ($\sigma=1,2,3$) in $d$-dimensional space. Any two-body interactions between particles are assumed to be suppressed and only the three-body potential \cite{Valiente_19} is switched on between fermions of different sorts. In order to take into account the finite-range effects of the three-body interaction, we consider a two-channel model that is very similar to that of the narrow Feshbach resonance but in the three-body sector
\begin{eqnarray}\label{H}
H=\sum_{\sigma}\int_{\bf p}\varepsilon_{\sigma}({\bf p})f^{\dagger}_{\sigma,{\bf p}}f_{\sigma,{\bf p}}+\int_{\bf p}\left[\frac{{\bf p}^2}{2M}+\delta\nu_{\Lambda}\right]c^{\dagger}_{\bf p} c_{\bf p}\nonumber\\
	+g\int_{{\bf p}_{1},{\bf p}_{2},{\bf p}_{3}}c^{\dagger}_{{\bf p}_{1}+{\bf p}_{2}+{\bf p}_{3}}f_{3,{\bf p}_{3}}f_{2,{\bf p}_{2}}f_{1,{\bf p}_{1}}+\textrm{h.c.},
\end{eqnarray}
where $\varepsilon_{\sigma}({\bf p})=\frac{{\bf p}^2}{2m_{\sigma}}$, $\int_{\bf p}=\int  \frac{d{\bf p}}{(2\pi)^d}$ and $M=m_1+m_2+m_3$ is the mass of composite fermion; $g$ and $\delta\nu_{\Lambda}$ are the coupling and the detuning. The latter depends on the ultraviolet (UV) cutoff $\Lambda$ that restricts the upper integration limit in the above momentum integrals. The anti-commutators of the field operators are fixed as follows $\{c_{\bf p}, c^{\dagger}_{{\bf p}'}\}=(2\pi)^d\delta({\bf p}-{\bf p}')$, $\{f_{\sigma,{\bf p}}, f^{\dagger}_{\sigma',{\bf p}'}\}=(2\pi)^d\delta_{\sigma,\sigma'}\delta({\bf p}-{\bf p}')$ and zero all other pairs. Before we proceed with the four-body bound and scattering states, let us first consider the simplest interacting system of three fermions. The latter solution sets up the correct renormalization of the cutoff-dependent detuning $\delta\nu_{\Lambda}$ and the relation of our model to the system with three-body $\delta$-like interaction \cite{Hryhorchak_22}. The three-body state that describes particles with zero total momentum can be written down as follows:
\begin{align}\label{c_state}
&|c\rangle=Ac^{\dagger}_{\bf 0}|0\rangle\nonumber\\
&+\int_{{\bf p}_{1},{\bf p}_{2}}A_{{\bf p}_{1},{\bf p}_{2},-{\bf p}_{1}-{\bf p}_{2}}f^{\dagger}_{1,{\bf p}_1}f^{\dagger}_{2,{\bf p}_2}f^{\dagger}_{3,-{\bf p}_{1}-{\bf p}_{2}}|0\rangle,
\end{align}
where $|0\rangle$ is the vacuum state, and amplitudes $A$ and $A_{{\bf p}_{1},{\bf p}_{2},{\bf p}_{3}}$ are subject to the Schr\"odinger equation. Denoting energy of the system by $\mathcal{E}$, we have
\begin{align}\label{c_Eqs}
&[\delta\nu_{\Lambda}-\mathcal{E}]A+g\int_{{\bf p}_{1},{\bf p}_{2}}A_{{\bf p}_{1},{\bf p}_{2},-{\bf p}_{1}-{\bf p}_{2}}=0,\\
&\left[\sum_{\sigma}\varepsilon_{\sigma}({\bf p}_{\sigma})-\mathcal{E}\right]A_{{\bf p}_{1},{\bf p}_{2},{\bf p}_{3}}+gA=0.
\end{align}
The non-trivial solution to these coupled equations for the bound states $\epsilon_g$ (negative $\mathcal{E}$s) reads
\begin{align}
\mathcal{D}(\epsilon_g)=0,
\end{align}
where for latter convenience we have introduced an auxiliary function
\begin{align}\label{D_E}
\mathcal{D}(\mathcal{E})=\delta\nu_{\Lambda}-\mathcal{E}
-g^2\int_{{\bf p}}\Pi_{23}({\bf p}|\mathcal{E}),
\end{align}
here and below $\Pi_{23}({\bf p}|\mathcal{E})=\Pi_{23}\left(\mathcal{E}-\frac{{\bf p}^2}{2M_{23}}-\varepsilon_{1}({\bf p})\right)$, $\Pi_{23}({\bf p},{\bf p}'|\mathcal{E})=\Pi_{23}\left(\mathcal{E}-\frac{({\bf p}+{\bf p}')^2}{2M_{23}}-\varepsilon_{1}({\bf p})-\varepsilon_{1}({\bf p}')\right), \dots$ with $M_{23}=m_2+m_3$ and $\Pi_{23}(\mathcal{E})=\int_{{\bf p}}\frac{1}{\varepsilon_{2}({\bf p})+\varepsilon_{3}({\bf p})-\mathcal{E}}$. First, let us discuss the limit of broad resonance. Keeping energy fixed while setting  $g\to \infty$, we can identify the three-body bare coupling constant $g^{-1}_{3,\Lambda}=-\delta\nu_{\Lambda}/g^2$ which absorbs the UV divergence of the integral in $\mathcal{D}(\mathcal{E})$. This procedure relates the observable three-body coupling $g_3$ to the three-body bound state energy $\epsilon_{\infty}$ at the broad resonance
\begin{eqnarray}\label{g_3}
g^{-1}_{3}=-\frac{\Gamma(1-d)}{(2\pi)^d}\left(\frac{m_1m_2m_3}{M}\right)^{d/2}|\epsilon_{\infty}|^{d-1}.
\end{eqnarray}
Restoring the finite magnitude of $g$ and repeating the above calculation procedure, one obtains the transcendental equation for the composite fermion bound state energy at narrow resonance 
\begin{eqnarray}\label{e_g}
\epsilon_{g}+\frac{g^2}{g_3}\left[1-\left(\frac{\epsilon_g}{\epsilon_{\infty}}\right)^{d-1}\right]=0.
\end{eqnarray}
There is always a single solution to this equation for positive $g_3$s. At large coupling $g$ it recovers the broad-resonant result $\epsilon_{\infty}$, while in a case of small $g$s asymptotically behaves as $\epsilon_{g}\approx-g^2/g_3$. The unitarity limit, $\epsilon_{g}=0$, is reached at infinite $g_3$ for all dimensions above $d=1$. The one-dimensional limit of Eq.~\ref{e_g} is also well-defined
\begin{eqnarray}\label{e_g_d_1}
\epsilon_{g}=\frac{g^2}{2\pi}\sqrt{\frac{m_1m_2m_3}{M}}\ln\left(\frac{\epsilon_g}{\epsilon_{\infty}}\right),
\end{eqnarray}
although for the three-body coupling (\ref{g_3}) is not. From Eq.~(\ref{e_g_d_1}) it can be readily concluded that $|\epsilon_{g}|\le|\epsilon_{\infty}|$. Finally, the Galilean invariance allows us to generalize the above results to a non-zero composite fermion momentum ${\bf p}$: the only modification is the energy shift $\mathcal{E}\to\mathcal{E}-\frac{{\bf p}^2}{2M}$ in Eq.~(\ref{D_E}).

\section{Four-body problem}
The wave function of an arbitrary four-body (composite fermion$+f_1$-atom) state with zero center-of-mass momentum reads
\begin{align}\label{fc_state}
&|f_1c\rangle=\int_{{\bf p}}B_{{\bf p}}f^{\dagger}_{1,{\bf p}}c^{\dagger}_{-{\bf p}}|0\rangle
+\int_{{\bf p}_{1},{\bf p}'_{1},{\bf p}_{2}}B_{{\bf p}_{1}{\bf p}'_1,{\bf p}_{2},-{\bf p}_{1}-{\bf p}'_1-{\bf p}_{2}}\nonumber\\
&\times f^{\dagger}_{1,{\bf p}_1}f^{\dagger}_{1,{\bf p}'_1}f^{\dagger}_{2,{\bf p}_2}f^{\dagger}_{3,-{\bf p}_{1}-{\bf p}'_1-{\bf p}_{2}}|0\rangle,
\end{align}
where $B_{{\bf p}_{1}{\bf p}'_1,{\bf p}_{2},-{\bf p}_{1}-{\bf p}'_1-{\bf p}_{2}}=-B_{{\bf p}'_{1}{\bf p}_1,{\bf p}_{2},-{\bf p}_{1}-{\bf p}'_1-{\bf p}_{2}}$ is anti-symmetric function of first two arguments. By acting with the Hamiltonian $H$ on the ansatz (\ref{fc_state}) one obtains the system of two coupled equations
\begin{align}
\label{fc_Eq_1}
&\left[\varepsilon_1({\bf p})+\frac{{\bf p}^2}{2M}+\delta\nu_{\Lambda}-\mathcal{E}\right]B_{\bf p}\nonumber\\
&+2g\int_{{\bf p}_{1},{\bf p}_{2}}B_{{\bf p}{\bf p}_1,{\bf p}_{2},-{\bf p}-{\bf p}_1-{\bf p}_{2}}=0,\\
&\label{fc_Eq_2}\left[\sum_{\sigma}\varepsilon_{\sigma}({\bf p}_{\sigma})+\varepsilon_1({\bf p}'_1)-\mathcal{E}\right]B_{{\bf p}_{1}{\bf p}'_1,{\bf p}_{2},{\bf p}_{3}}\nonumber\\
&+\frac{g}{2}\left[B_{{\bf p}_1}-B_{{\bf p}'_1}\right]=0.
\end{align}
for the wave-function with eigenvalue $\mathcal{E}$. For negative energies $\mathcal{E}=\epsilon_4$ (four-body bound states) Eqs.~(\ref{fc_Eq_1}), (\ref{fc_Eq_2}) can be readily transformed into a single integral equation for the function $B_{{\bf p}}$
\begin{align}\label{B_p}
\mathcal{D}_{1}({\bf p}|\epsilon_4)
B_{\bf p}
+g^2\int_{{\bf q}}\Pi_{23}\left({\bf p},{\bf q}|\epsilon_4\right)B_{{\bf q}}=0,
\end{align}
with the shorthand notation for function $\mathcal{D}_{1}({\bf p}|\mathcal{E})=\mathcal{D}\left(\mathcal{E}-\varepsilon_1({\bf p})-\frac{{\bf p}^2}{2M}\right)$. Because of the Fermi statistics, the four-body bound states can potentially occur only in the odd-wave channels. The one requiring the shallowest potential well is the $p$-wave with the wave function of the following form $B_{\bf p}=({\bf n}{\bf p}/p)B_p$ (here $B_p$ depends on modulus $\bf p$ with ${\bf n}$ being a unit vector). As it is expected for states with the orbital quantum number $l=1$, this is not a single wave function, but $d$ mutually-orthogonal states parameterized by different ${\bf n}_i$ satisfying the condition ${\bf n}_i{\bf n}_j=\delta_{ij}$. Plugging the $p$-wave harmonics back into Eq.~(\ref{B_p}) and performing the $d$-dimensional hyper-angle integration, we obtained the one-dimensional integral equation (see Appendix) for amplitude $B_p$. 

\subsection{$d>1$}
\begin{figure}[h!]
	\centerline{\includegraphics
		[width=0.45
		\textwidth,clip,angle=-0]{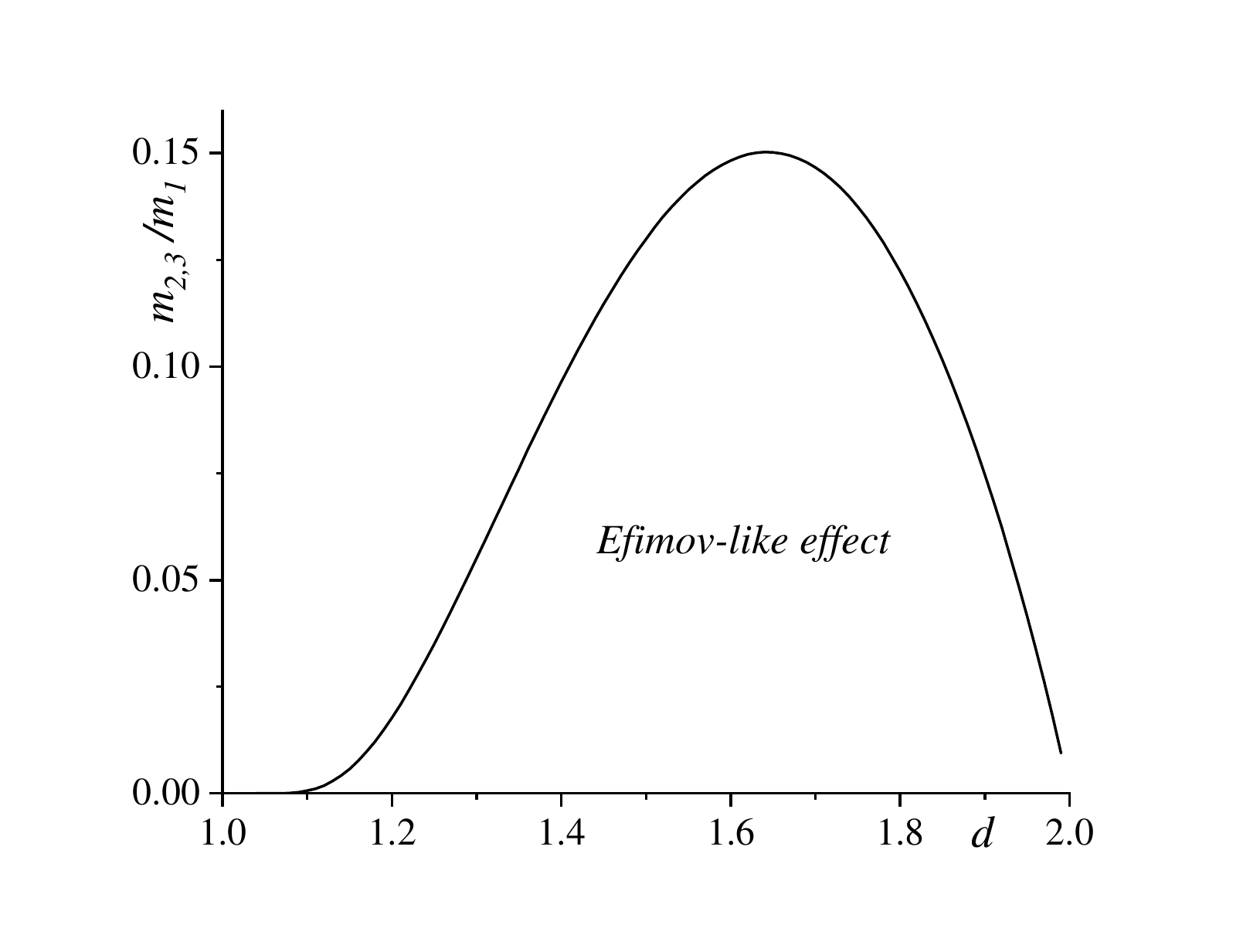}}
	\caption{Region, where the $p$-wave Efimov-like effect in the four-body sector of the three-component Fermi system with the three-body interaction emerges.}\label{Efimov_fig}
\end{figure}

Before discussing numerical results, let us consider the limit of broad resonance $g\to \infty$ and disappearing three-body bound state $g_3\to \infty$. In the case of bosons, this limit is characterized by the universal log-periodic wave function that signals an emergence \cite{Hryhorchak_22} of the Efimov-like effect in the four-body sector. A very similar situation should be observed in our system. However, its fermionic nature (and, consequently, non-zero total angular momentum) demands the total mass of the second and third particles $M_{23}$ should be small enough in comparison to $m_1$ in order to provide the effective attracting potential for trapping of four particles. The scaling limit ($g,g_3\to \infty$) supposes the power-law behavior $B_p=1/p^{1-d/2\pm \eta}$ of the wave function. Then, the integral equation transforms into the algebraic one on parameter $\eta$
\begin{eqnarray}\label{algebraic_Eq}
1=\frac{\Gamma\left(1-\frac{d/2+\eta}{2}\right)\Gamma\left(1-\frac{d/2-\eta}{2}\right)m_1M^{2d-3}}{-\Gamma(1-d)\Gamma(1+d/2)M^{d-1}_{23}(M+m_1)^{d-1}}\nonumber\\
\times _{2}F_{1}\left(1-\frac{d/2+\eta}{2},1-\frac{d/2-\eta}{2};\frac{2+d}{2};\frac{m^2_1}{M^2}\right),
\end{eqnarray}
with $\Gamma(z)$ and $_{2}F_{1}\left(a,b;c;z\right)$ being the gamma and hypergeometric functions \cite{Abramowitz}, respectively. Truly imaginary solutions $\eta=i\eta_0$ determine a region (see Fig.~\ref{Efimov_fig})
of the four-body $p$-wave Efimov-like effect in the considered system, with the universal ratio of energy levels
\begin{eqnarray}\label{eps_ratio}
\epsilon^{(n)}_4/\epsilon^{(n+1)}_4=e^{2\pi/\eta_0}, \ \ (n\gg 1).
\end{eqnarray}
Note that $\eta_0$ depends only on the spatial dimension and mass ratio $m_1/M_{23}$. The asymptotic form of the appropriate wave functions is a linear combination of two partial solutions
\begin{eqnarray}\label{B_p_as}
B_p=\frac{\sin\left(\eta_0\ln (p/\Lambda_0)\right)}{p^{1-d/2}},
\end{eqnarray}
where $\Lambda_0$ is an arbitrary momentum scale related to the lowest Efimov state. An emergent discrete scale invariance of the asymptotic expression (\ref{B_p_as}) for the wave function in the scaling region is the characteristic feature of the Efimov-like effects.

\begin{figure}[h!]
	\centerline{\includegraphics
		[width=0.5
		\textwidth,clip,angle=-0]{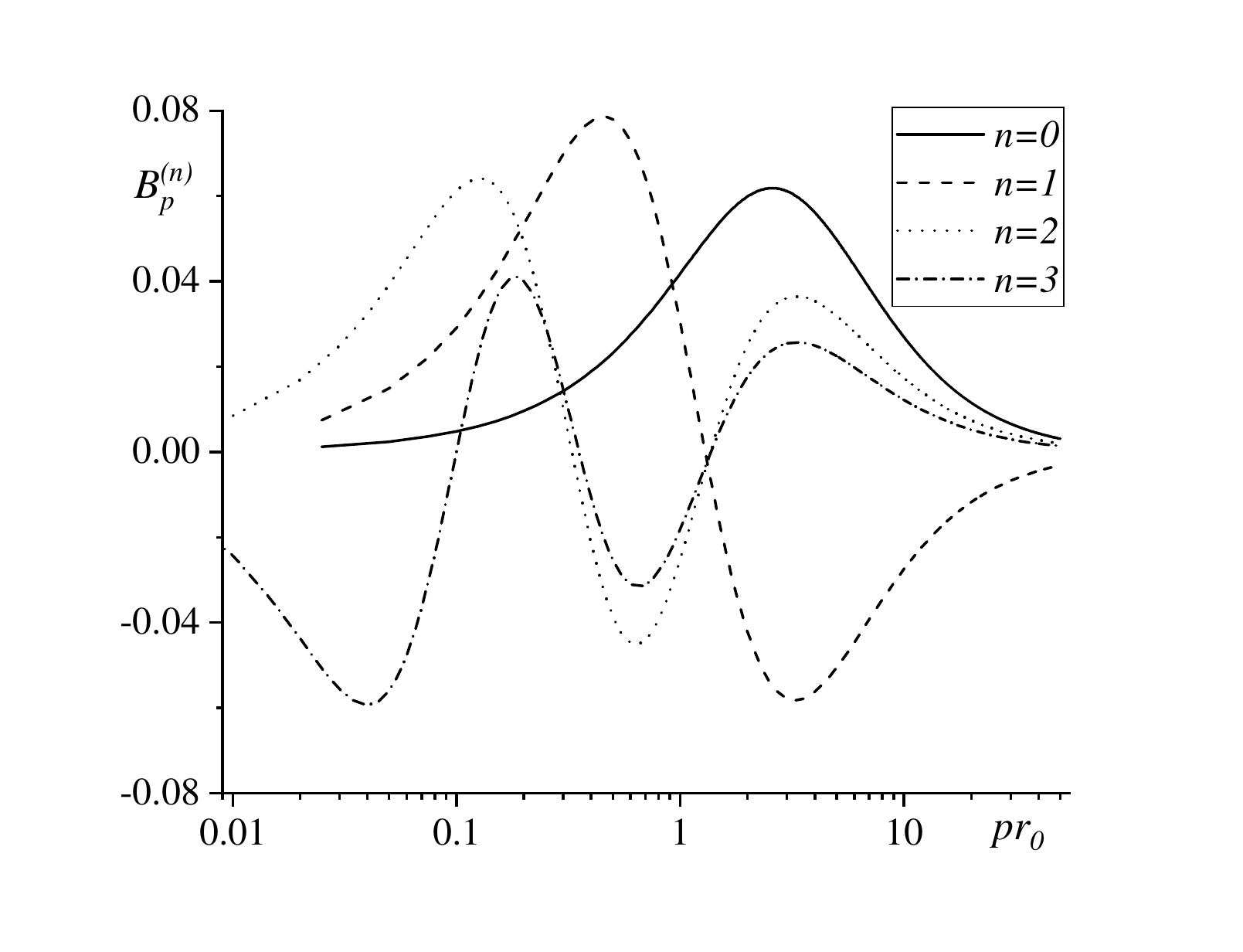}}
	\caption{The ground state and the first few excited states wave functions $B^{(n)}_p$ (unnormalized) of the four-body problem in the $p$-wave channel.}\label{wave_functions_fig}
\end{figure}

To verify the above predictions about the behavior of our model in the four-particle sector, we have numerically solved the integral equation (\ref{B_p}) in the $p$-wave channel by discretizing its kernel. The parameters of the system were specially chosen to reveal the Efimov behavior as simple as possible. In particular, we put $d=1.5$ and mass ratios $M_{23}/m_1=1/50$ small enough to provide a strong induced attractive potential between two $f_1$-atoms. The three-body binding energy is sent to zero $\epsilon_g\to 0$, and coupling $g=\sqrt{\frac{m_1m_2m_3}{M}}\frac{1}{r^{2-d}_0}$ is parametrized by scale $r_0$ which is related to the effective range. Actually, this is the only dimensionful parameter at unitary limit $g^{-1}_3=0$. The four-body $p$-wave energies  $\epsilon^{(n)}_4$ are measured in units of $\frac{m_1+M}{2m_1Mr^2_0}$, and the numerical prefactors for the first five levels are gathered in table~\ref{tab:table1}. 
\begin{table}[h!]
	\begin{center}
		\begin{tabular}{c|c|c} 
			$n$ & $|\epsilon^{(n)}_4|$ & $\epsilon^{(n-1)}_4/\epsilon^{(n)}_4$ \\
			\hline
			0  & 4.2250    & \\
			1  & 3.4802$\times10^{-1}$  & 12.141\\
			2  & 3.5671$\times10^{-2}$  & 9.7564\\
			3  & 3.8086$\times10^{-3}$  & 9.3668\\
		    4  & 4.0949$\times10^{-4}$  & 9.3008\\			
	  $\vdots$ & $\vdots$               & $\vdots$\\
	  $\infty$ & 0                      & 9.2909\\
		\end{tabular}
	\caption{The first five eigenvalues $\epsilon^{(n)}_4$ (in units of $\frac{m_1+M}{2m_1Mr^2_0}$) of the four-body problem in the $p$-wave channel. Numerical calculations were performed in $d=1.5$ and for mass ratios $m_2=m_3=m_1/100$.}\label{tab:table1}
	\end{center}
\end{table}
For the identification of the level, the wave functions were also calculated (see Fig.~\ref{wave_functions_fig}).
In the third column, we show the ratio of neighboring eigenenergies (\ref{eps_ratio}) which should tend to $e^{2\pi/\eta_0}=9.290926...$ at large $n$. It is seen, however, that already for the fourth excited state this quantity $\epsilon^{(3)}_4/\epsilon^{(4)}_4$ deviates from the universal value by a tenth of a percent.

\subsection{$d=1$}
\begin{figure}[h!]
\includegraphics[width=0.23\textwidth]{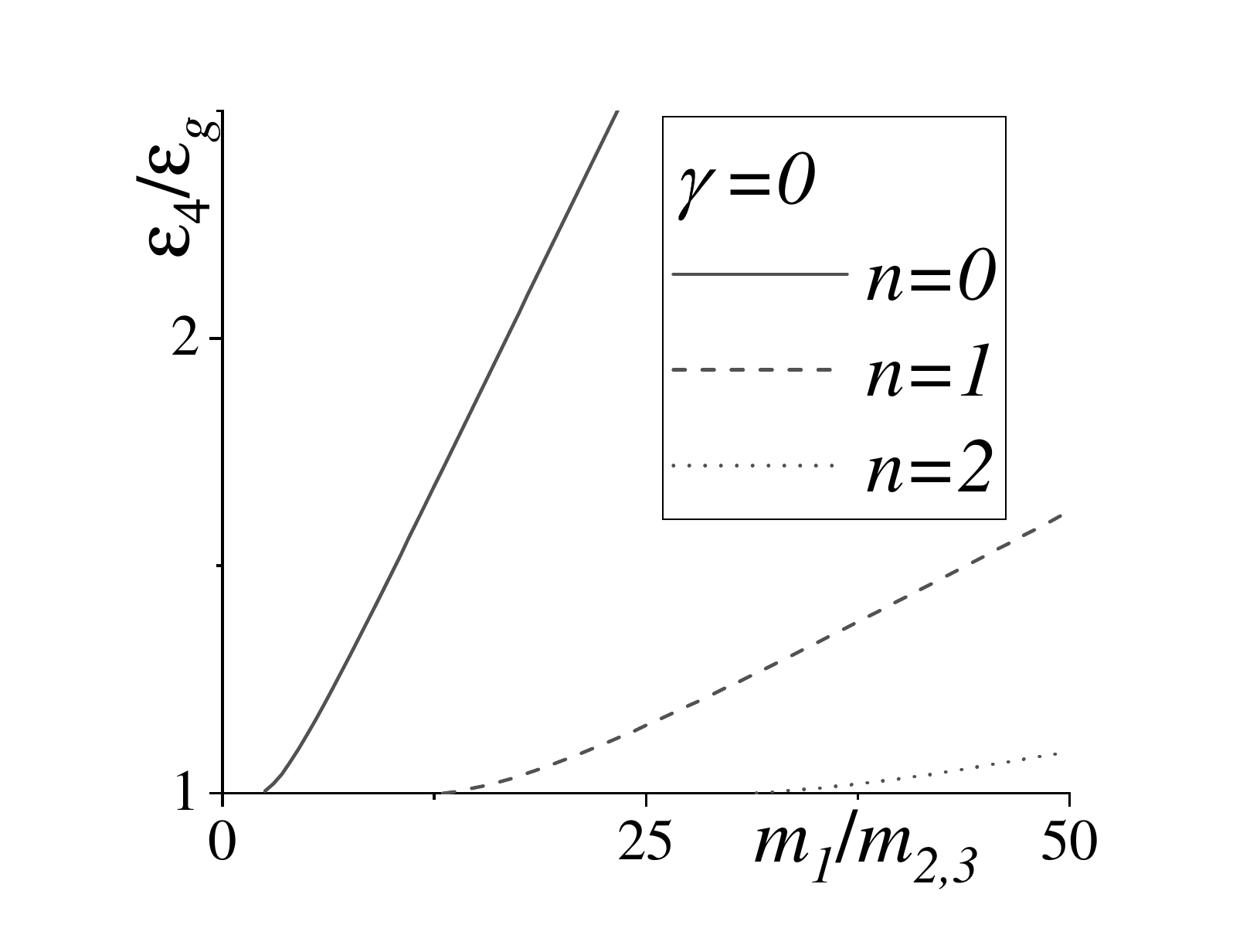}
\includegraphics[width=0.23\textwidth]{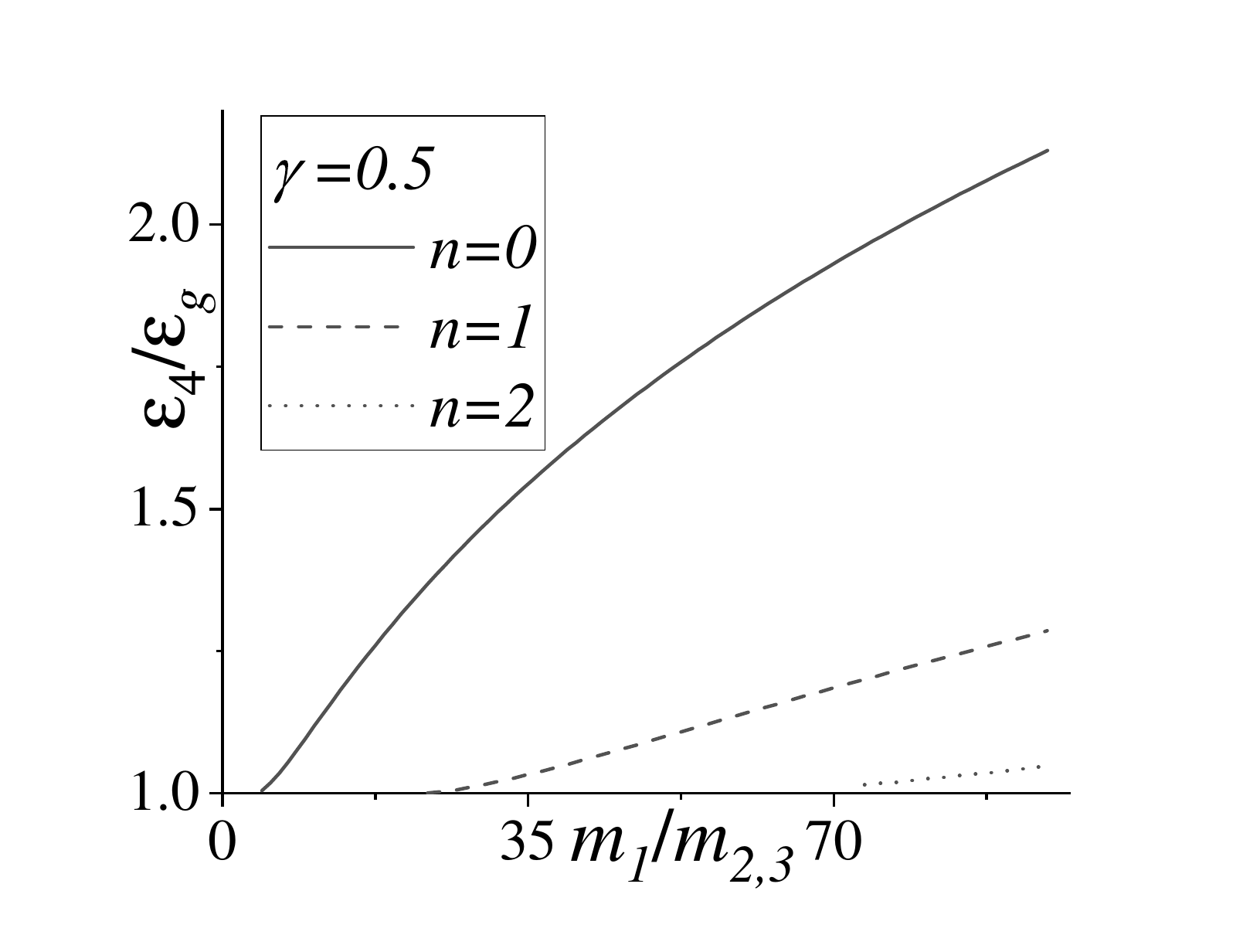}
\includegraphics[width=0.23\textwidth]{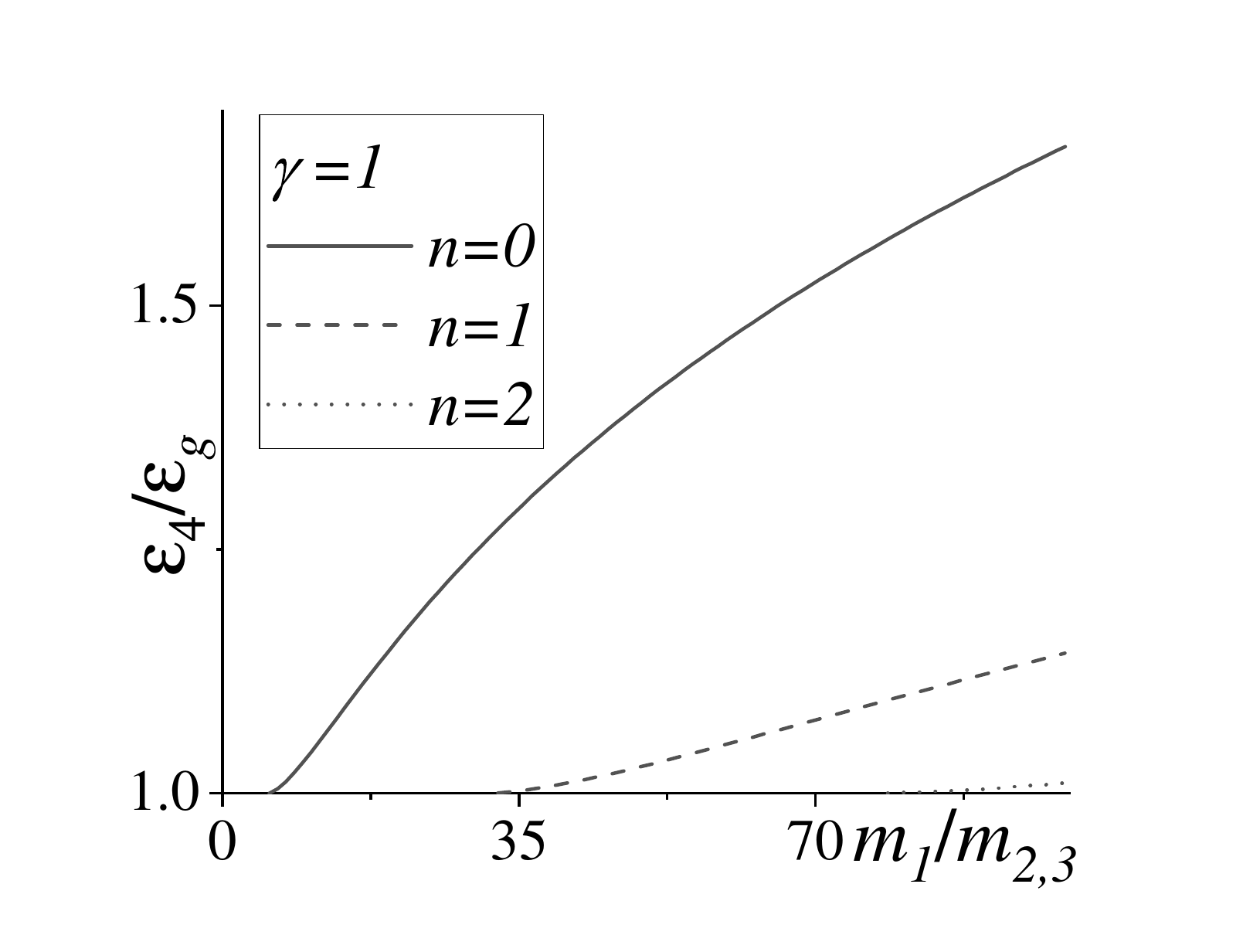}
\includegraphics[width=0.23\textwidth]{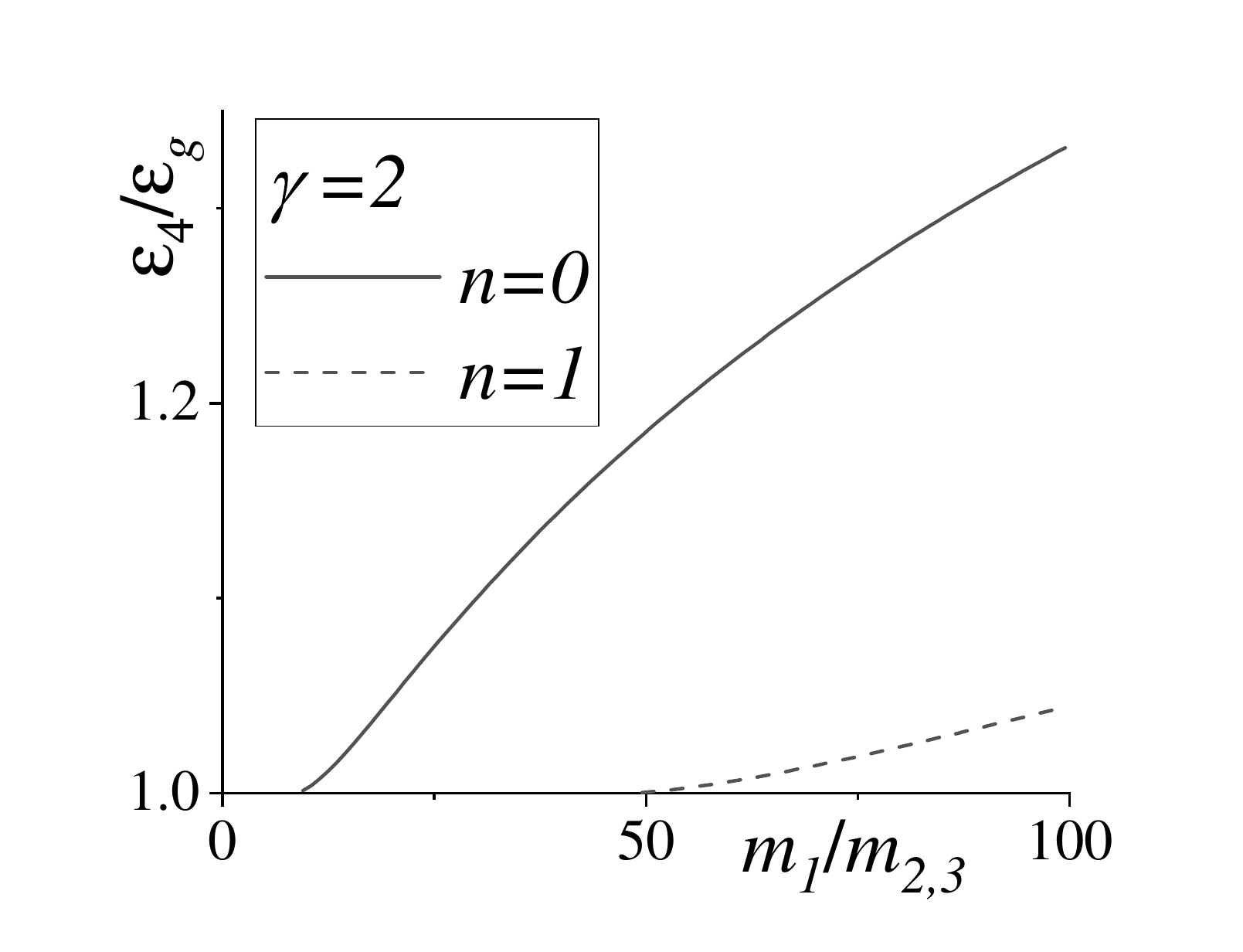}
\caption{Four-body energy levels (in units of $\epsilon_g$) as functions of mass ratio $m_1/m_{2,3}$ at the different effective ranges (parameterized by $\gamma=\ln\left(\frac{\epsilon_{\infty}}{\epsilon_g}\right)$) of the three-body interaction.}\label{eigenvalues_fig}
\end{figure}

\begin{figure}[h!]
	\centerline{\includegraphics
		[width=0.45
		\textwidth,clip,angle=-0]{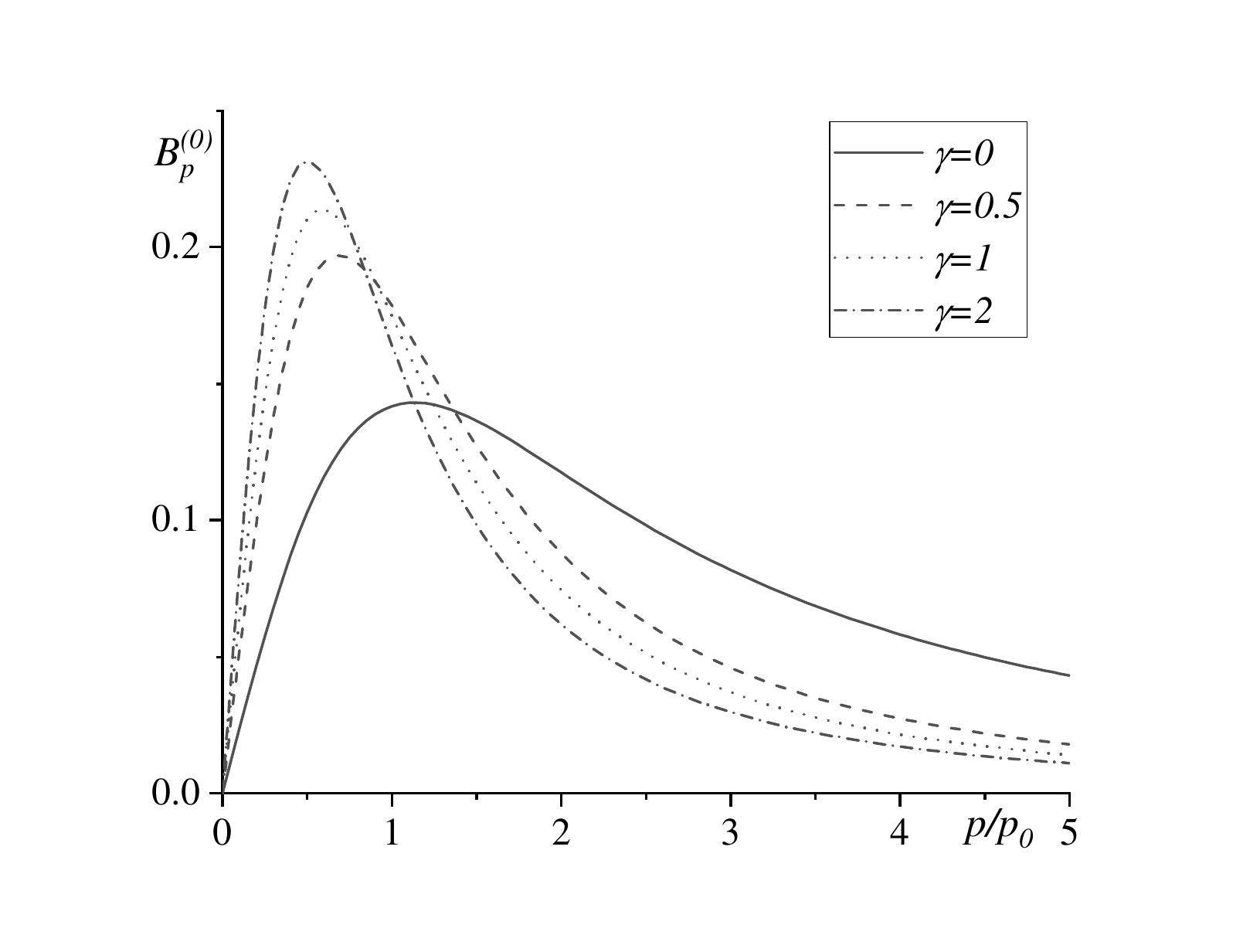}}
	\caption{The four-body ground state wave functions $B^{(0)}_p$ (unnormalized) at different values of the effective range and fixed mass ratios $m_1/m_{2}=m_1/m_{3}=15.5$ in 1D.}\label{1D_wave_functions_fig}
\end{figure}

\begin{figure}[h!]
	\includegraphics[width=0.235\textwidth]{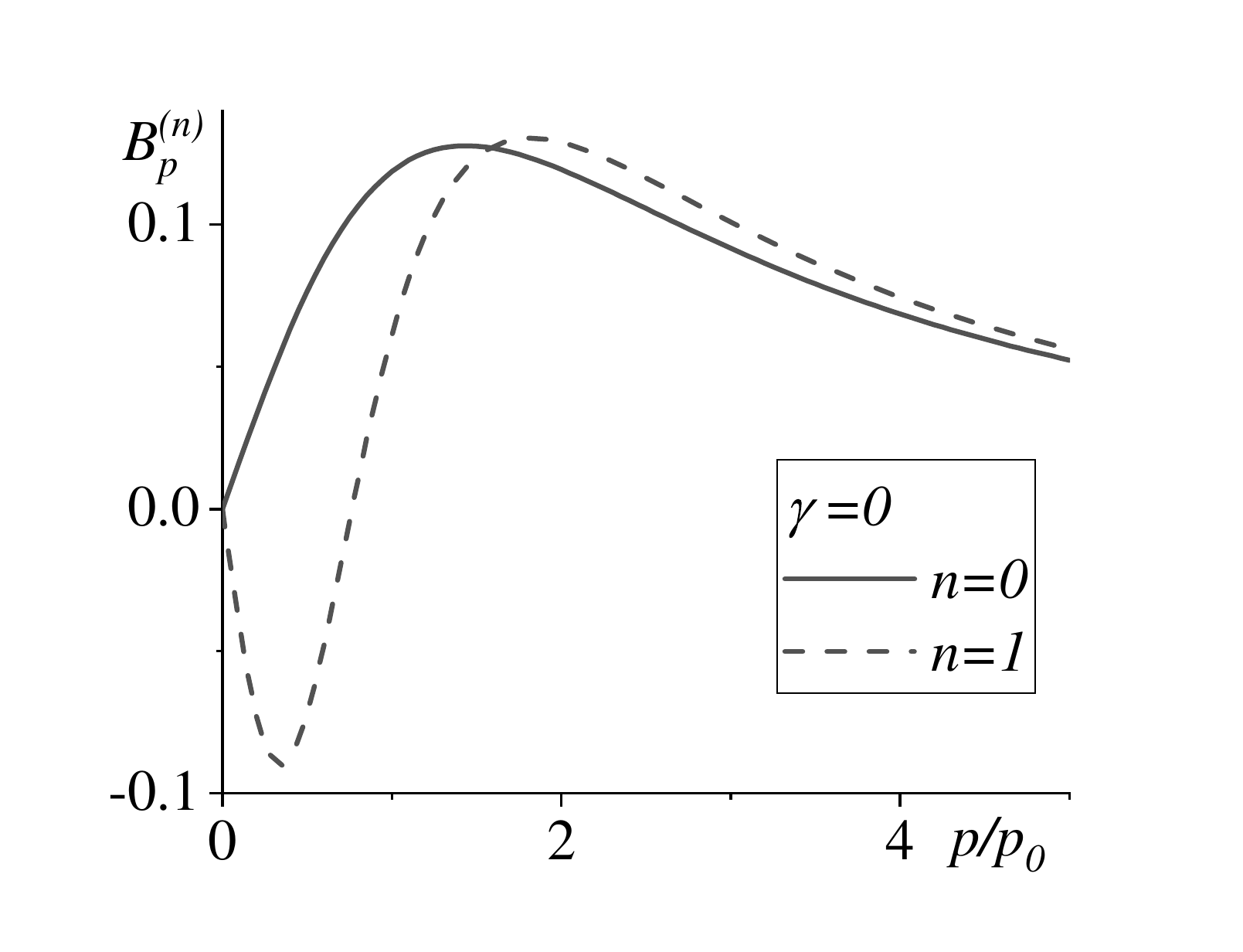}
	\includegraphics[width=0.235\textwidth]{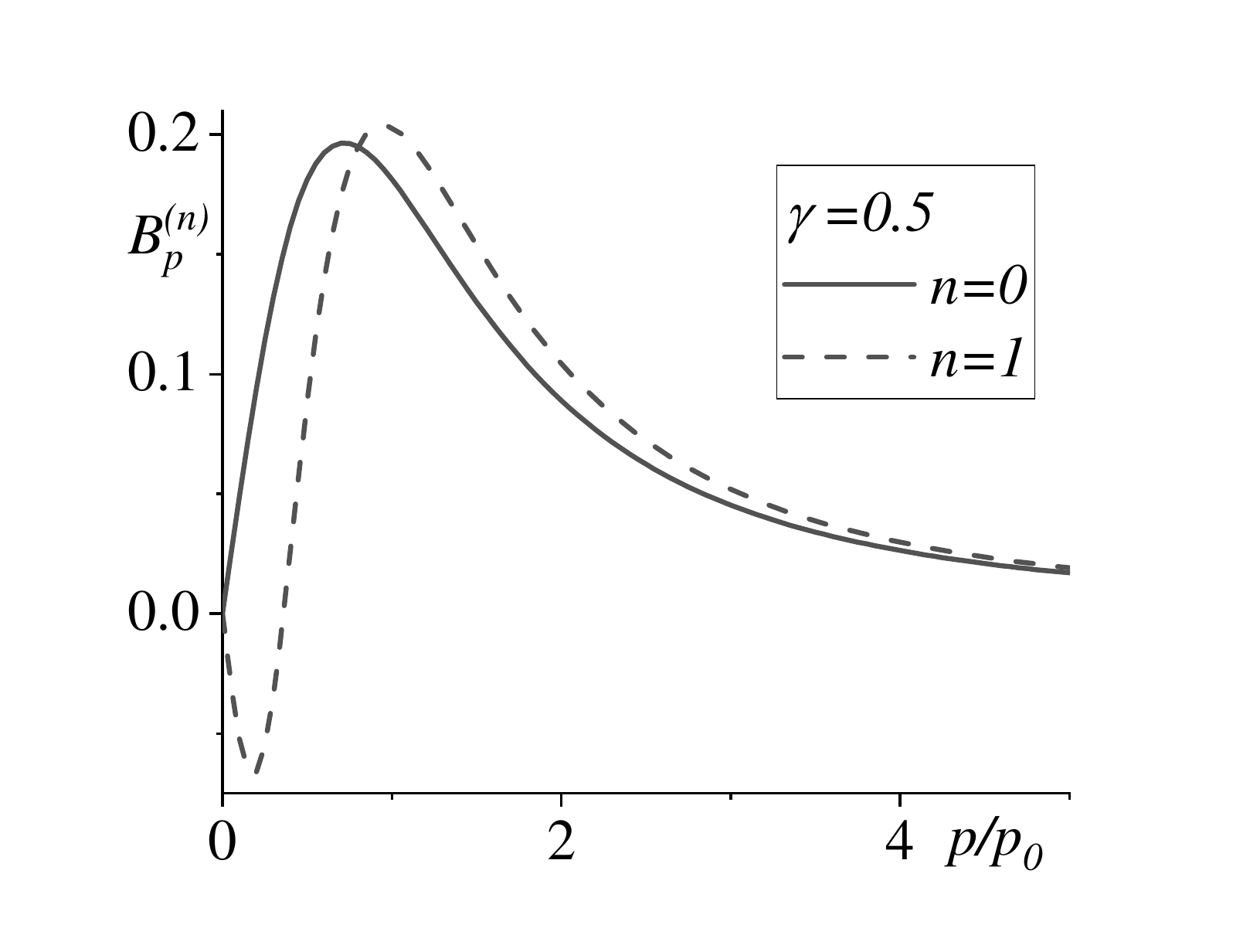}
	\caption{Four-body wave functions (unnormalized) of the first two energy levels for mass ratio $m_1/m_{2,3}=30.5$ at two effective ranges $\gamma=0$ (left) and $\gamma=0.5$ (right).}\label{wf_fig}
\end{figure}

The one-dimensional case is of particular interest due to potential realization in experiments and/or numerical simulations. According to the previous section, there is no four-body Efimov-like effect in the considered system in 1D, but a strong imbalance between masses of atoms may lead to the formation of the bound states. Projection onto $p$-wave states in the four-body Shr\"odinger equation~(\ref{B_p}) in $d=1$ reduces to choosing the odd $B_{\bf p}=-B_{-{\bf p}}$ solutions. An appropriate eigenvalues $\epsilon_4$ at finite ranges ($g<\infty$) of the interaction potential, should be deeper than the three-body bound-states $\epsilon_g$. Therefore, it is natural to pick dimensionless units in a way that the four-body energies are measured in units of $\epsilon_g$, and the effective range is determined by closeness of the three-body bound state energy to its broad-resonance value $\gamma=\ln\left(\frac{\epsilon_{\infty}}{\epsilon_g}\right)$. Then $\gamma=0$ corresponds to zero range $r_0=0$ case, while $\gamma>0$ refers to the model with narrow resonance. Introducing momentum scale $p_0$ such that $\frac{m_1+M}{2m_1M}p^2_0=|\epsilon_g|$, one can readily rewrite Eq.~{\ref{B_p}} in dimensionless units in 1D 
\begin{align}\label{b_p}
&\int^{\infty}_{-\infty} dq\frac{b_q
}{\sqrt{\left(q-\frac{m_1}{M}p\right)^2+\left(1-\frac{m^2_1}{M^2}\right)(p^2+\lambda)}}\nonumber\\
&=\left[\gamma(p^2+\lambda-1)+\ln(p^2+\lambda)\right]
b_{p},
\end{align}
with $\lambda=\epsilon_4/\epsilon_g$ being dimensionless eigenvalue and $B_{p_0p}=b_{p}$. Recall that due to the fermionic nature of the considered system, solutions for $b_{p}$ should be searched on a class of odd functions. The results of numerical diagonalization are presented in Fig.~\ref{eigenvalues_fig}.
Since, both $m_2$ and $m_3$ enters Eq.~(\ref{b_p}) only in combination $M_{23}$, we can freely take them equal to each other $m_2=m_3$. Our calculations demonstrate that even at the broad resonance $\gamma=0$, which is the most favorable limit for the existence of the four-body bound states, the first level emerges at $m_1/m_2\approx 2.5$ (the second and the third ones at $\approx 13.0$ and $\approx 31.5$, respectively). With an increase of the effective range, these values shift towards larger $m_1/m_2$ ratios. Particularly, when $\epsilon_g=\epsilon_{\infty}/e$, the first three tetramers are found at $m_1/m_2\approx 5.5$, $m_1/m_2\approx 32.5$ and $m_1/m_2\approx 78.5$. We have also obtained the four-body ground-state wave functions in 1D (see Fig.~\ref{1D_wave_functions_fig}) with $m_1/m_{2}=15.5$
at various effective ranges. For comparison, the first two four-body eigenstates are depicted in Fig.~\ref{wf_fig} 
for zero $\gamma=0$ and non-zero $\gamma=0.5$ effective ranges of the three-body interaction.

Figure~\ref{eigenvalues_fig} (especially zero-range case) reveals an intrinsic dependence of the eigenvalues at the large mass imbalance. Indeed, integral in Eq.~\ref{b_p} possesses a non-integrable singularity at $q=\frac{m_1}{M}p$ when $m_1/M =1$. This fact together with parity properties of the wave function $b_p$ allows for the asymptotic (with logarithmic precision) determination of the four-body bound states
\begin{align}\label{eps_4_as}
\gamma(\lambda-1)+\ln\lambda-\ln\frac{C_{\gamma}}{1-m^2_1/M^2}+\dots=0.
\end{align}
Constant $C_{\gamma}$ should be determined for every four-body energy level separately by imposing the solution to Eq.~(\ref{eps_4_as}) to be consistent with the large-$\frac{m_1}{M_{23}}$ tail of the eigenvalue behavior found numerically. From Eq.~(\ref{eps_4_as}) one immediately recognizes the linear dependence of $\epsilon_4$ on $m_1/m_{2,3}$ (in the limit $m_1/m_{2,3}\gg 1$) at broad resonance. Although the asymptotic solution (\ref{eps_4_as}) qualitatively explains the eigenvalue behavior at large mass imbalance, the logarithmic accuracy does not allow the quantitative description of the numerically calculated curves in Fig.~\ref{eigenvalues_fig} at intermediate mass ratios.

\section{Summary}
In conclusion, we have proposed for the first time the effective model of contact three-body interaction that includes the effects of finiteness of the potential range. Applying this effective description to the three-component Fermi system with the suppressed two-body interactions in the fractional dimension above $d=1$, we have predicted the emergence of the Efimov-like physics in the $p$-wave channel of the four-body sector. An analytic estimations in the scaling limit are supported by numerically exact calculations for finite effective ranges at unitarity. The detailed analysis of the one-dimensional problem revealed the necessary conditions for the occurrence of negative eigenvalues in the four-body spectrum both in the case of broad and narrow resonances. Particularly, it is shown that depending on mass ratios of fermions with the three-body interaction, one can in principle observe an arbitrarily large number of the tetramer levels. The effect is suppressed for the non-zero effective ranges towards larger mass imbalance.

\begin{center}
	{\bf Acknowledgements}
\end{center}
Discussions with O.~Hryhorchak on the related problem are gratefully acknowledged. This work was partly supported by Project No.~0122U001514 from the Ministry of Education and Science of Ukraine.

\section{Appendix}
For self-sufficiency of the material presented, all important explicit expressions for functions used in the main text are collected in this section. In arbitrary $d$ the $f_2-f_3$ vacuum bubble reads
\begin{eqnarray}
\Pi_{23}(\mathcal{E})=\frac{\Gamma(1-d/2)}{(2\pi)^{d/2}}\left(\frac{m_2m_3}{M_{23}}\right)^{d/2}(-\mathcal{E})^{d/2-1}.
\end{eqnarray}
The $p$-wave partial harmonics of function $\Pi_{23}\left({\bf p},{\bf q}|\epsilon_4\right)$ in Eq.~\ref{B_p} is found after the solid-angle averaging
\begin{eqnarray}
\frac{1}{\Omega_d}\int d\Omega_d\frac{{\bf p}{\bf q}}{pq}\Pi_{23}\left({\bf p},{\bf q}|\epsilon_4\right)=\pi_{23}(p,q|\epsilon_4)\nonumber\\
=-\frac{\Gamma(2-d/2)}{(2\pi)^{d/2}d}\left(\frac{m_2m_3}{M_{23}}\right)^{d/2}\nonumber\\
\times\frac{z\,_2F_1(1-\frac{d}{4},\frac{3}{2}-\frac{d}{4};1+\frac{d}{2};z^2)}{\left(\frac{p^2+q^2}{2m_1M_{23}/M}+|\epsilon_4|\right)^{1-d/2}},
\end{eqnarray}
where $z=\frac{pq/M_{23}}{\frac{p^2+q^2}{2m_1M_{23}/M}+|\epsilon_4|}$. Then the final one-dimensional integral equation utilized for numerical calculations in arbitrary dimension $d$ reads
\begin{eqnarray}
\mathcal{D}_{1}({\bf p}|\epsilon_4)B_p
=\frac{\Omega_d}{(2\pi)^d}\int^{\infty}_0dqq^{d-1}\pi_{23}(p,q|\epsilon_4)B_q,
\end{eqnarray}
with function in the r.h.s. explicitly given by 
$\mathcal{D}_{1}({\bf p}|\epsilon_4)=\frac{p^2}{2m_1}+\frac{p^2}{2M}+|\epsilon_{4}|+\frac{g^2}{g_3}\left[\frac{\left(\frac{p^2}{2m_1}+\frac{p^2}{2M}+|\epsilon_{4}|\right)^{d-1}}{|\epsilon_{\infty}|^{d-1}}-1\right]$. In $d=1$ case the above equation reduces, after passing to dimensionless variables, to Eq.~(\ref{b_p}) in the main text.

\end{document}